\documentclass[conference]{IEEEtran}
\IEEEoverridecommandlockouts

\usepackage{cite}
\usepackage{amsmath,amssymb,amsfonts}
\usepackage{algorithm}
\usepackage{algpseudocode}
\usepackage{graphicx}
\usepackage{textcomp}
\usepackage{xcolor}
\usepackage{makecell}
\usepackage{enumitem}
\usepackage{subcaption}
\usepackage{pifont}
\usepackage{url}
\usepackage{hyperref}
\usepackage{lineno}

\def\BibTeX{{\rm B\kern-.05em{\sc i\kern-.025em b}\kern-.08em T\kern-.1667em\lower.7ex\hbox{E}\kern-.125emX}}

\newif\ifshowdiff
\ifshowdiff
  \newcommand{\new}[1]{\textcolor{red}{#1}}
  \newcommand{\gfix}[1]{\textcolor{blue}{#1}}   % language / grammar corrections
  \newcommand{\gnew}[1]{\textcolor{green!55!black}{#1}}  % text updated to match new Figs. 7-8
  \newcommand{\final}[1]{\textcolor{magenta}{#1}}       % final pre-submission pass
  \newcommand{\fillin}[1]{\textcolor{blue}{\textbf{[FILL IN: #1]}}}
\else
  \newcommand{\new}[1]{#1}
  \newcommand{\gfix}[1]{#1}
  \newcommand{\gnew}[1]{#1}
  \newcommand{\final}[1]{#1}
  \newcommand{\fillin}[1]{}
\fi

\begin{document}

% \linenumbers
% \setlength\linenumbersep{2pt} 

\title{HDLouvain: High-Throughput Incremental\\Clustering for Hyperdimensional Data}

\author{%
\IEEEauthorblockN{Beenish Gul\textsuperscript{1},
                  Mahbod Afarin\textsuperscript{2},
                  Farhat Ahmed\textsuperscript{2},
                  Sumukh Pinge\textsuperscript{2},
                  Tsung-Han Lu\textsuperscript{2},
                  Tajana Rosing\textsuperscript{2},
                  Kevin Skadron\textsuperscript{1}}
\IEEEauthorblockA{\textsuperscript{1}\textit{University of Virginia}, Charlottesville, USA \qquad
                  \textsuperscript{2}\textit{University of California, San Diego}, San Diego, USA\\
\{bg9qq, skadron\}@virginia.edu \qquad
\{mafarin, f3ahmed, spinge, tsl012, tajana\}@ucsd.edu}
}

\maketitle

\begin{abstract}
Modern shotgun proteomics generates tens of thousands of tandem mass spectra per
hour, producing hundreds of gigabytes of acquisition data that strain
conventional clustering pipelines. \new{Existing approaches face a fundamental
trade-off. DBSCAN and hierarchical clustering (HAC) enable flexible
density-based and multi-resolution similarity search respectively, but DBSCAN is
unstable across varying data densities, HAC costs $O(n^2)$, and both yield poor
cluster quality. Graph-based methods such as Louvain produce high-quality
clusters but scale poorly on streaming data. We observe that Louvain's
clustering quality and its poor scalability are not fundamentally at odds, and
that the latter can be resolved without sacrificing the former.} This
observation motivates a framework that integrates Louvain's high-fidelity
community detection with the lightweight, bit-parallel similarity computation of
hyperdimensional computing (HDC). HDLouvain achieves this by partitioning an
initial bootstrap fraction of the stream into precursor-mass buckets and
applying Louvain to each bucket's $k$-NN similarity graph, condensing every
resulting community into a single binary hypervector through bitwise-majority
bundling rather than the iterative centroid recomputation used by $k$-means or
HAC. New data is then assigned through a fast, GPU-accelerated Hamming-distance
lookup, matching each spectrum against representatives in its mass bucket or
seeding a new cluster when no match meets the similarity threshold $\tau$,
allowing the partition to grow incrementally without re-clustering the full
dataset. \new{We evaluate HDLouvain on five PRIDE datasets spanning 0.99\,M to
21.13\,M spectra, sweeping the bootstrap fraction $\alpha$ from 0.2 to 0.9}, and
find that it exposes an explicit and monotonic throughput--fidelity trade-off
rather than a single operating point. \new{On PXD003258, the largest dataset in
the runtime sweep, speedup ranges from 4.5$\times$ at $\alpha = 0.3$ to
1.65$\times$ at $\alpha = 0.7$. Across the swept datasets, modularity rises
monotonically from 0.75--0.76 at $\alpha = 0.2$ to 0.95--0.97 at $\alpha = 0.9$,
and never falls more than 8\% below full static recomputation.}
\final{Independently of $\alpha$, the Louvain community-detection stage scores
0.86--0.93 normalized mutual information (NMI) on LFR benchmark networks with
planted ground truth from 30k to 2M nodes,} \new{whereas DBSCAN and HAC degrade
to 0.57.}

\end{abstract}


\section{Introduction}
\label{intro}

\gfix{Mass spectrometry (MS)-based proteomics} has become indispensable for drug discovery, biomarker identification, and personalized medicine~\cite{ponzina2025smartms}. Modern mass spectrometers produce tens of thousands of tandem mass spectra per hour, and public repositories such as MassIVE and PRIDE now host hundreds of terabytes of raw spectral data, with submission rates growing by an order of magnitude over the past decade~\cite{deutsch2020proteomexchange, perezriverol2019pride, ucsd2024massive, pinge2024rapidoms}. Spectral clustering is the standard strategy for taming this redundancy. It groups highly similar spectra together and represents each group with a single consensus spectrum. Clustering shrinks the dataset, speeds up database search, and lets unidentified spectra be identified by association with similar spectra in the same group~\cite{wu2023spectroscape}. However, as data generation outpaces available compute resources, clustering itself has emerged as the primary bottleneck.

\new{This is a big-data problem rather than a \gfix{modeling} one, since the algorithmic ingredients are standard and what breaks at repository scale is data management. Volume rules out exact clustering, as the dense distance matrix over the 21\,million spectra of a single PRIDE submission holds $4.46\times10^{14}$ entries, some 446\,TB at one byte each. Velocity is even more problematic, because a method whose per-spectrum cost grows with the volume already ingested accumulates $\Theta(n^{2})$ work against a $\Theta(n)$ time budget and so falls irrecoverably behind acquisition. Sustained ingestion demands a per-item cost independent of cumulative volume.}

\noindent
\textit{\textbf{Challenge 1} -- The scalability-quality trade-off in static clustering}. Existing MS clustering tools occupy opposite ends of a quality-speed spectrum. Graph-based and hierarchical methods such as MaRaCluster~\cite{the2016maracluster} and hierarchical agglomerative clustering (HAC)~\cite{murtagh2011hac} cluster well but scale poorly, since exhaustive all-pairs comparison is $O(n^2)$ and even accelerated variants take hours to days at repository scale~\cite{wang2019mscrush, xu2023hyperspec}; approximate methods such as msCRUSH~\cite{wang2019mscrush} and Falcon~\cite{bittremieux2021falcon} trade fidelity for runtime. Hyperdimensional computing (HDC) offers a middle ground, encoding each spectrum as a binary hypervector so that similarity reduces to a bit-parallel XOR and popcount, which sustains high throughput on GPUs and FPGAs~\cite{pinge2024spechd, xu2023hyperspec}. Its known weakness is that cluster quality depends heavily on prototype initialization, and random seeding yields unstable partitions~\cite{ge2024robust}. In short, fast HDC methods lack a principled way to \emph{discover} clusters, while high-quality graph methods cannot keep pace with streaming data.

\begin{figure*}
\centering
\includegraphics[scale=0.42]{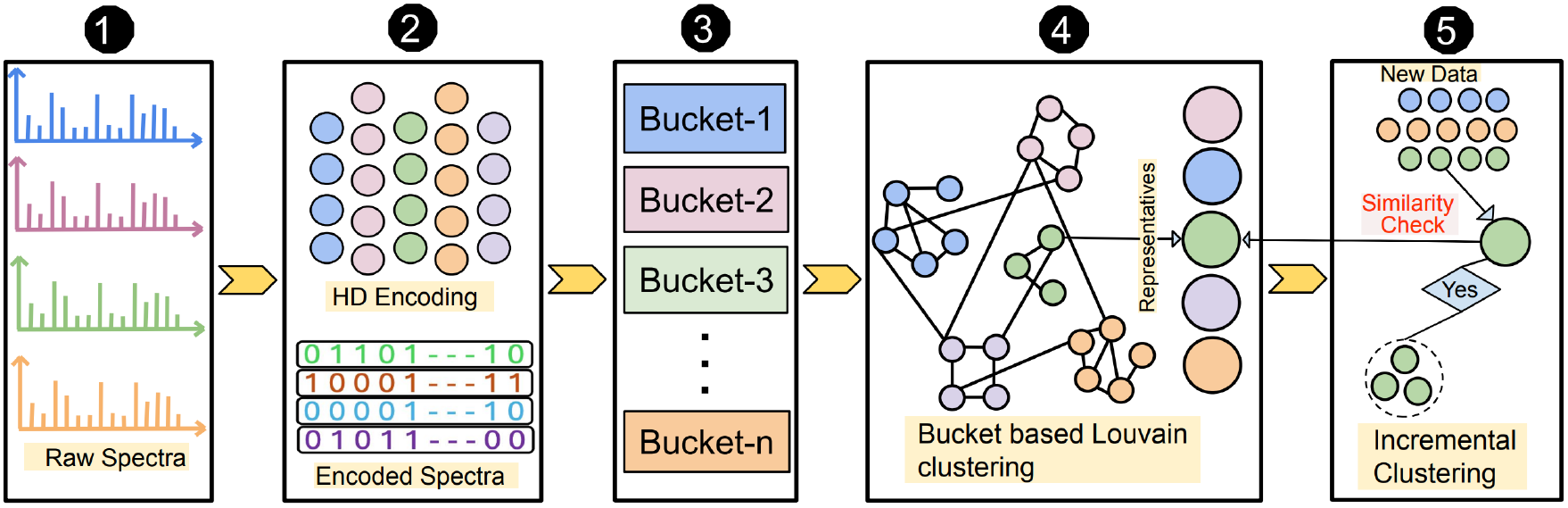}
% \vspace{-0.1in}
      \caption{\textbf{The HDLouvain pipeline.} \textit{Offline (\ding{182}--\ding{185}):} spectra are HD-encoded into binary hypervectors, bucketed by precursor mass, and clustered per bucket by Louvain on a $k$-NN graph; \new{each community is then bundled once into one representative hypervector.} \textit{Online (\ding{186}):} \new{each new spectrum is bucketed by Eq.~\eqref{eq:bucket}, encoded, and matched by XOR/popcount against that bucket's representatives only}, or seeds a new cluster when no match is found.}
\label{fig:overview}
\vspace{-0.15in}
\end{figure*}

\noindent
\textit{\textbf{Challenge 2} -- The absence of incremental clustering for continuous acquisition}. A second, largely unaddressed limitation is that every scalable HDC-based MS tool to date (HyperSpec~\cite{xu2023hyperspec}, SpecHD~\cite{pinge2024spechd}, RapidOMS~\cite{pinge2024rapidoms}, and SmartMS~\cite{ponzina2025smartms}) operates in a static, batch-oriented mode that requires the entire dataset up front, whereas spectra arrive continuously from the instrument and re-clustering from scratch on every batch is prohibitively expensive. The classical alternatives do not help, since DBSCAN~\cite{ester1996dbscan} and HAC~\cite{murtagh2011hac} were designed for static datasets and their naive online adaptations drift in quality as clusters grow. Louvain is likewise a batch algorithm, because modularity optimization assumes a fixed graph and sweeps repeatedly over all nodes. Incorporating new spectra therefore means either rebuilding the $k$-NN graph and re-running Louvain at full offline cost, or using dynamic variants that hold quality only under modest perturbations and still recompute per batch. Neither sustains the arrival rate of a running instrument.

We propose \textbf{HDLouvain}, a hybrid incremental clustering framework that decouples cluster \emph{discovery} from cluster \emph{assignment}. Discovery is a single upfront investment made with Louvain~\cite{blondel2008louvain}, a modularity heuristic that runs in near-linear time and needs neither a predefined cluster count nor density parameters~\cite{sahu2024dflouvain,hernandez2025sna}. HDLouvain treats the offline/online split as a tunable parameter $\alpha$ rather than a fixed design choice, and we evaluate $\alpha \in \{0.3, \ldots, 0.7\}$ \final{for runtime and $\alpha$ from 0.2 to 0.9 for clustering quality} to characterize the trade-off between upfront discovery cost and online assignment quality. \new{Each discovered community is then bundled into a single binary representative hypervector in one pass at the end of the offline phase. This is the only prototype computation HDLouvain performs; unlike $k$-means or HAC-based pipelines~\cite{to2021clustersheep}, the representative is never recomputed as the stream grows,} \final{so per-spectrum online cost is set by the bucket's representative set rather than by the volume already ingested.}

\textcolor{black}{Fig.~\ref{fig:overview}} summarizes the pipeline as five stages across an offline and an online phase. Offline, spectra (\ding{182}) are encoded into binary hypervectors by ID--Level HD encoding (\ding{183}), sorted by precursor $m/z$, and partitioned into buckets (\ding{184}) that confine pairwise comparisons to mass-compatible candidates and avoid the global $O(n^2)$ cost; within each bucket a $k$-NN graph is built, Louvain recovers high-modularity communities (\ding{185}), \new{and each is condensed into one representative hypervector.} Online (\ding{186}), \new{each new spectrum is mapped to its bucket by Eq.~\eqref{eq:bucket}, encoded, and compared against the representatives \emph{of that bucket only}} by Hamming distance, joining the closest one within threshold $\tau$ or else seeding a new cluster. Novel spectra, \gfix{for example,} peptides absent from the offline partition, are therefore absorbed without a re-run of Louvain, and because assignment touches only bucket-local representatives, per-spectrum cost scales with the size of that representative set rather than with the dataset.

\vspace{0.1in}
\noindent In summary, our contributions are as follows:

\begin{itemize}[leftmargin=*]
    \item \textbf{A hybrid algorithm that unifies community detection and HDC.} To our knowledge, HDLouvain is the first framework to seed an HDC clustering pipeline for mass spectrometry with Louvain community detection. \new{Louvain supplies the cluster \emph{membership} HDC lacks a principled way to discover, and bundling turns it into a bit-parallel prototype in a single pass over the bootstrap data.}
 \item \textbf{An incremental clustering design for streaming spectra.} A bucket-local scheme confines Louvain to a one-time offline phase and performs all online assignment by GPU XOR/popcount against cluster representatives, with bit-packing cutting per-spectrum cost 32-fold and no full re-clustering.
 \item \new{\textbf{A characterization of the throughput--fidelity trade-off.} On five MS proteomics datasets spanning 0.99\,M to 21.13\,M spectra, HDLouvain reaches $3.21\times$ over static Louvain at 50/50 bootstrap/incremental }\final{ and $4.5\times$ at 30/70 on the four evaluated datasets}\new{, with the bootstrap fraction  $\alpha$ acting as a single monotonic knob on the trade-off.} \final{Its Louvain stage scores 0.86--0.93 NMI (Normalized Mutual Information) on benchmarks,} \new{while DBSCAN and HAC collapse to 0.57 beyond 30k nodes and fall three orders of magnitude below Louvain under ARI (Adjusted Rand Index).} \gnew{Measured on the MS data rather than only on synthetic graphs, modularity climbs
from 0.75--0.76 at a 20\% bootstrap to 0.95--0.97 at 90\% and never falls more than 8\% below full static recomputation, bounding the
fidelity cost of the decomposition.}
\end{itemize}

\section{Background and Motivation}

\subsection{Mass Spectrometry Clustering}

Shotgun proteomics generates tens of thousands of tandem mass spectra per hour~\cite{wu2023spectroscape, deutsch2020proteomexchange}, and public repositories now hold tens of billions~\cite{perezriverol2019pride}. \new{A tandem (MS/MS) spectrum is a list of fragment peaks, each a mass-to-charge ratio ($m/z$) paired with an intensity, together with the $m/z$ and charge of the \emph{precursor} ion that was fragmented to produce it. The precursor $m/z$ therefore acts as a coarse, instrument-supplied identifier of the underlying peptide mass, and we exploit it in Section~\ref{sec:bucketing}.} Clustering tames the resulting redundancy by grouping spectra of the same peptide, so that downstream analysis such as database searching treats each cluster as one representative spectrum. Beyond compression this improves reusability, since unidentified spectra can be identified by association with an identified neighbor, identification errors corrected, and modified peptides or outliers detected through cluster patterns~\cite{wu2023spectroscape}.

Clustering at this scale is itself expensive, since finding nearest neighbors in principle requires comparing every spectrum against every other~\cite{wu2023spectroscape}, and state-of-the-art tools still take hours to days, negating the speed gains of acquisition~\cite{xu2023hyperspec}. Accelerated designs such as the GPU-based ClusterSheep~\cite{to2021clustersheep} and the hashing and indexing approaches of msCRUSH~\cite{wang2019mscrush} and Falcon~\cite{bittremieux2021falcon} cut the comparison count but remain batch procedures, which motivates incremental clustering in a setting where acquisition never stops and repositories grow continuously~\cite{wang2026incremental}.

\subsection{Louvain Algorithm}
\label{sec:bg-louvain}

Louvain performs unsupervised clustering as community detection on graphs. Developed by Blondel et al.~\cite{blondel2008louvain}, it is widely used for its efficiency, scalability, and result quality~\cite{hernandez2025sna}. It greedily optimizes modularity, a measure of connection density inside communities versus outside, merging nodes into whichever community yields the greatest gain~\cite{sahu2024dflouvain}, and runs essentially linearly in the number of edges. Modularity suits our setting because it needs no ground-truth labels, which real spectral data lacks, and rewards a partition only when its communities are denser than a degree-preserving null model predicts, so it measures whether recovered clusters are structurally real rather than arbitrary; where labels exist, on the LFR benchmarks of Section~\ref{sec:accuracy}, we also report NMI and ARI. \new{Louvain is a \emph{heuristic}, not an exact solver. Modularity maximization is NP-hard~\cite{brandes2008modularity}, so it returns a high-modularity partition with no optimality guarantee, and different node orderings can yield different partitions.} Leiden~\cite{traag2019leiden} repairs its tendency to produce internally disconnected communities, but adds passes at every level, whereas HDLouvain runs community detection once, offline on the bootstrap portion, over small per-bucket graphs where that pathology is bounded by bucket occupancy; nothing in the design depends on this choice, and Leiden can be substituted directly. Applied to a similarity graph whose nodes are spectra and whose edges join similar pairs, Louvain yields densely interconnected clusters with which to initialize incremental clustering.

%\beenish{We retain Louvain because Leiden's extra refinement phase is worth its computational overhead only when applied to the full dataset, where its stronger connectivity guarantees can be fully realized. In our setting, however, Louvain is applied to just a bootstrap fraction of the stream, so the online phase cannot claim full clustering accuracy regardless of the algorithm used, since subsequent data is clustered based on representative hypervectors rather than being re-clustered in full. Given this, Louvain's output is already comparably accurate to what Leiden would offer at this scale, making Leiden's added computational cost difficult to justify for only marginal accuracy gains. Louvain therefore provides the better balance of accuracy and runtime for this setting, making it the most suitable choice for our real-time streaming workload}

\subsection{Hyperdimensional Computing}
\label{sec:bg-hdc}

Hyperdimensional Computing (HDC) represents information as very high-dimensional binary vectors called hypervectors (HVs), manipulated by simple operations such as bitwise XOR, bundling, and permutation~\cite{pinge2024rapidoms, pinge2024spechd}. Because these operations are low-precision and bit-parallel across thousands of dimensions, HDC is efficient on modern hardware and noise-tolerant through its distributed representation~\cite{heddes2024hdc, kanerva2009hdc}.

Two operations matter here. \textbf{Bundling} aggregates a set of HVs into one composite vector. \new{For binary hypervectors it is a \emph{bitwise majority}, so the $\ell$-th bit is 1 exactly when more than half of the inputs have a 1 in position $\ell$. The result is closer, in Hamming distance, to every member of the set than a random hypervector would be, which is what makes it usable as a cluster representative.} The \textbf{similarity metric} is \new{the normalized Hamming distance between two binary HVs $\mathbf{h}_i$ and $\mathbf{h}_j$,}

\begin{equation}
d(\mathbf{h}_i, \mathbf{h}_j) = \frac{\text{popcount}(\mathbf{h}_i \oplus \mathbf{h}_j)}{D} \;\in\; [0,1],
\label{eq:hamming}
\end{equation}

\new{where $\oplus$ is bitwise XOR and $\text{popcount}(\cdot)$ counts set bits. Eq.~\eqref{eq:hamming} is the only distance HDLouvain uses, applied unchanged in Sections~\ref{sec:knn} and~\ref{sec:online}.} \new{We do not claim binary HDC beats the quantized representations now common in machine learning; it is an extreme point of that design space, and what the extreme point buys is operational. At one bit per dimension a similarity evaluation is a fixed-width integer instruction with no floating-point unit and no dequantization step, so 32 dimensions are compared per word and a bucket's representative set fits in on-chip memory,} making HDC a good match for a streaming pipeline.

\section{HDLouvain Design}
\label{sec:design}

HDLouvain integrates the modularity-driven community detection of the Louvain algorithm with the computational efficiency of HDC to enable scalable clustering over high-dimensional spectral data. \new{Algorithm~\ref{alg:hdlouvain} states the complete method, and the remainder of this section describes each stage together with the reasoning behind it. The framework is governed by four algorithmic parameters\gfix{, namely the bootstrap fraction $\alpha$, the neighborhood size $k$, the hypervector dimension $D$, and the assignment threshold $\tau$}, together with the spectrum-preprocessing settings given in Section~\ref{sec:preprocessing}. Spectrum parsing and precursor-mass bucketing run on the CPU, while HD encoding, \gfix{pairwise distance computation, and} online assignment run on the GPU.}

\begin{algorithm}[t]
\caption{\new{HDLouvain}}
\label{alg:hdlouvain}
\small
\begin{algorithmic}[1]
\Require stream $S$; bootstrap fraction $\alpha$; \gfix{neighborhood} size $k$; HV dimension $D$; threshold $\tau$
\Ensure a partition of $S$ into disjoint clusters\gfix{, namely the offline communities $\bigcup_b \mathcal{C}^{*}_b$ extended by the online assignments}
\Statex \emph{Convention:} $\mathcal{R}_b = \emptyset$ for every bucket $b$ \gfix{that the offline phase does not initialise}, and $\min_{r \in \emptyset} d(\cdot,r) = \infty$.
\Statex \rule{\linewidth}{0.4pt}
\Statex \textbf{Offline phase} \emph{(once, on the bootstrap prefix)}
\State $S_{\text{off}} \gets$ first $\lfloor \alpha |S| \rfloor$ spectra of $S$
\ForAll{$s \in S_{\text{off}}$} \Comment{data-parallel}
  \State $b(s) \gets$ bucket index of $s$ by Eq.~\eqref{eq:bucket} \Comment{$O(1)$}
  \State $h_s \gets \textsc{Encode}(\textsc{Preprocess}(s))$ \Comment{$h_s \in \{0,1\}^D$}
\EndFor
\State sort $S_{\text{off}}$ by precursor $m/z$; let $V_b = \{s : b(s) = b\}$, $n_b = |V_b|$ \Comment{$O(|S_{\text{off}}| \log |S_{\text{off}}|)$}
\ForAll{buckets $b$ with $V_b \neq \emptyset$}
  \State $\mathcal{R}_b \gets \emptyset$ \Comment{representatives of $b$}
  \State $\Delta^{(b)}_{ij} \gets d(h_i, h_j)$, $\ \forall\, v_i, v_j \in V_b$ \Comment{$O(n_b^2 D/32)$}
  \State $N_k(v_i) \gets \textsc{Select}_k\big(\gfix{\{\Delta^{(b)}_{ij}\}_{j \neq i}}\big)$, $\ \forall\, v_i \in V_b$ \Comment{$O(n_b^2)$}
  \State $E_b \gets \big\{ (i,j) : v_j \in N_k(v_i) \ \textbf{or} \ v_i \in N_k(v_j) \big\}$
  \State $W_b \gets \big\{ w_{ij} = 1 - \Delta^{(b)}_{ij} : (i,j) \in E_b \big\}$ \Comment{Eq.~\eqref{eq:weight}}
  \State $\mathcal{C}^{*}_b \gets \textsc{Louvain}\big(G_b = (V_b, E_b, W_b)\big)$ \Comment{Eq.~\eqref{eq:modularity}; $O(k\,n_b)$}
  \ForAll{$C \in \mathcal{C}^{*}_b$}
    \State $r_C \gets \textsc{Bundle}\big(\{h_s : s \in C\}\big)$ \Comment{Eq.~\eqref{eq:bundle}; $O(|C|D)$}
    \State $\mathcal{R}_b \gets \mathcal{R}_b \cup \{r_C\}$ \gfix{\Comment{$r_C$ identifies $C$}}
  \EndFor
\EndFor
\Statex \rule{\linewidth}{0.4pt}
\Statex \textbf{Online phase} \emph{(streaming, in arrival order)}
\For{each $s$ arriving after the bootstrap prefix}
  \State $b \gets b(s)$ by Eq.~\eqref{eq:bucket} \Comment{$O(1)$}
  \State $h_s \gets \textsc{Encode}(\textsc{Preprocess}(s))$
  \State $d^{\star} \gets \min_{r \in \mathcal{R}_b} d(h_s, r)$, attained at $r^{\star}$ \Comment{$O(|\mathcal{R}_b| D/32)$}
  \If{$d^{\star} \le \tau$} \Comment{false whenever $\mathcal{R}_b = \emptyset$}
    \State assign $s$ to the cluster of $r^{\star}$ \Comment{$r^{\star}$ \textbf{not} updated}
  \Else
    \State open a new cluster $C_{\text{new}} \gets \{s\}$ \gfix{identified by $h_s$}
    \State $\mathcal{R}_b \gets \mathcal{R}_b \cup \{h_s\}$ \Comment{$\textsc{Bundle}(\{h_s\}) = h_s$}
  \EndIf
\EndFor
\end{algorithmic}
\end{algorithm}

\subsection{Spectrum Preprocessing}
\label{sec:preprocessing}

HDLouvain accepts Mascot Generic Format (MGF) files as input\new{, a plain-text format in which each record holds the precursor $m/z$ and charge followed by a list of fragment $m/z$--intensity pairs,} and employs multiprocessing to read multiple files in parallel, distributing the workload across $k_{\text{cpu}}$ independent CPU cores.

\new{Preprocessing converts each parsed record into a fixed-length hypervector and consists of four steps applied independently per spectrum, which is what makes the stage embarrassingly parallel. First, peaks are filtered: the precursor peak and peaks outside the usable fragment $m/z$ window are discarded, and only the $N$ most intense remaining peaks are retained, which bounds the work per spectrum regardless of instrument settings. Second, intensities are square-root transformed and $L_2$-normalized, so that a small number of dominant peaks cannot alone determine similarity. Third, the surviving peaks are quantized onto a fixed $m/z$ grid of bin width $\delta_{m/z}$. Fourth, ID--Level encoding maps each occupied bin to a hypervector: a randomly drawn but fixed \emph{ID} hypervector identifies the bin, a \emph{level} hypervector encodes the quantized intensity, the two are bound with XOR, and the results are bundled across all occupied bins and binarized to yield $h_s \in \{0,1\}^D$. Nearby $m/z$ bins receive correlated level hypervectors, so small mass shifts produce small Hamming distances, which is the property the downstream graph construction relies on. We use $D = 2048$, $N = 50$, $\delta_{m/z} = 0.05$ and a fragment window of $[101~m/z,~1500~m/z]$ throughout.}

\subsection{Precursor-Mass Bucketing}
\label{sec:bucketing}

Once all MGF files have been preprocessed, the spectra are sorted in ascending order by precursor $m/z$ and assigned to buckets using Eq.~\eqref{eq:bucket}, as shown in \textcolor{black}{Fig.~\ref{fig:overview}}.

\begin{equation}
    \text{bucket}_i = \left\lfloor \frac{\left(\frac{m}{z_i} \cdot C_i\right) - 1.00794}{1.0005079} \right\rfloor
    \label{eq:bucket}
\end{equation}

Here, $C_i$ denotes the precursor charge of the $i$-th spectrum, and $m/z_i$ represents the precursor mass-to-charge ratio of the $i$-th spectrum. The constant $1.00794$ corresponds to the mass of a charge carrier (i.e., a proton), and $1.0005079$ represents the spacing between the centers of two adjacent clusters of physically possible peptide masses.

\new{Bucketing exists to avoid the dominant cost of large-scale spectral clustering, which is the $O(n^2)$ dense distance matrix induced by all-pairs comparison. Confining pairwise distances to spectra within the same bucket removes cross-bucket work entirely, and the resulting reduction in both memory footprint and computation is what makes the offline phase tractable at repository scale.}

\new{Two properties of Eq.~\eqref{eq:bucket} distinguish it from an arbitrary grid. First, it is not a uniform partition of the $m/z$ axis, because the divisor $1.0005079$ is the average spacing between the equidistant peptide-mass cluster \gfix{centers} that arise from the elemental composition of tryptic peptides~\cite{wolski2006peptidemass}, so bucket edges fall in the mass gaps that essentially no tryptic peptide occupies, rather than at arbitrary cut points. Two spectra of the same peptide therefore land in the same bucket unless their measured precursor masses straddle one of these physically improbable gaps. Second, the mapping is a constant-time arithmetic expression on values the instrument already reports, so it costs the same whether it is applied to the bootstrap set offline or to a single arriving spectrum online.}

\new{The scheme is nevertheless not boundary-free. Spectra whose precursor mass falls close to a bucket edge, \gfix{for instance,} because of instrument mass error or an isotope-assignment mistake, can be separated from otherwise identical spectra on the far side of the edge, which fragments a true cluster into two. The standard remedy is overlapping buckets, in which each spectrum is replicated into the \gfix{neighboring} bucket whenever it lies within a tolerance $\eta$ of an edge; this bounds the boundary loss at the cost of a $1 + O(\eta)$ increase in offline work and a post-hoc merge of duplicated assignments. HDLouvain uses non-overlapping buckets in the results reported here, and we regard quantifying the residual boundary loss, and the $\eta$ that removes it, as the most immediate open question raised by this design.}

\subsection{$k$-NN Graph Construction}
\label{sec:knn}

\new{Let $\Delta \in [0,1]^{|V| \times |V|}$ be the pairwise distance matrix of a bucket, where $\Delta_{ij} = d(h_i, h_j)$ is the normalized Hamming distance of Eq.~\eqref{eq:hamming} between spectra $v_i$ and $v_j$. We write the matrix as $\Delta$ rather than $D$ because $D$ denotes the hypervector dimension throughout this paper.} A $k$-nearest neighbor ($k$-NN) graph $G = (V, E, W)$ is constructed directly over the vertex set $V$.

\new{For each vertex $v_i \in V$, its $k$ nearest \gfix{neighbors} $N_k(v_i)$ are obtained by a partial selection over the $i$-th row of $\Delta$ rather than a full sort. Introselect returns the $k$ smallest entries in $O(|V|)$ time, against $O(|V| \log |V|)$ for sorting, and we never need the \gfix{neighbors} in ranked order, only the set.}

\new{The $k$-NN relation is not symmetric, since $v_j \in N_k(v_i)$ does not imply $v_i \in N_k(v_j)$, whereas the modularity objective of Eq.~\eqref{eq:modularity} is defined on an undirected graph. HDLouvain therefore symmetrizes by union: an edge $e_{ij} \in E$ exists if and only if $v_j \in N_k(v_i)$ \emph{or} $v_i \in N_k(v_j)$. The union rather than the intersection is the right choice here because a spectrum from a small, tight peptide cluster is often in the \gfix{neighbor} list of a hub spectrum without the reverse holding, and the mutual-$k$-NN variant would drop exactly those edges and fragment the cluster.}

\new{The edge weights require a distance-to-similarity transformation, which we state explicitly because the construction is not reproducible without it. Modularity, Eq.~\eqref{eq:modularity}, rewards \emph{high} intra-community edge weight, so the weights must encode similarity. Setting $w_{ij} = \Delta_{ij}$ would assign the largest weight to the most dissimilar pair and drive the optimizer to group the least similar spectra, inverting the objective. Because $\Delta_{ij}$ is already normalized to $[0,1]$ by Eq.~\eqref{eq:hamming}, its complement is a valid similarity on the same scale, and HDLouvain uses}
\begin{equation}
w_{ij} = 1 - \Delta_{ij}, \qquad w_{ij} \in [0,1],
\label{eq:weight}
\end{equation}
\new{so that a pair of identical hypervectors contributes weight $1$ and a pair at chance distance contributes approximately $0.5$. All modularity values, all reported partitions, and the results in Section~\ref{sec:eval} use this convention.} The resulting sparse $k$-NN graph encodes the local similarity structure among spectra within each bucket and serves as input to the community detection step.

\subsection{Community Detection}
\label{sec:communities}

The $k$-NN graph $G = (V, E, W)$ of each bucket is passed to the Louvain algorithm~\cite{blondel2008louvain}, which serves as the core clustering engine of HDLouvain. Louvain addresses several limitations of the alternatives. Hierarchical clustering suffers  quadratic time and space complexity. DBSCAN, which grows clusters outward from regions holding at least MinPts points within radius $\epsilon$, is highly sensitive to the choice those two hyperparameters, which are difficult to tune in high-dimensional spectral spaces. $K$-means fixes the number of clusters in advance alongside its assumption of spherical, equally sized clusters. \gfix{All of these properties make the alternatives impractical} in real-world spectral datasets where the number of distinct peptide groups and their geometric distribution are unknown.

\new{Louvain returns a partition $\mathcal{C}^{*} = \{C_1, \ldots, C_R\}$ of $V$ into disjoint, exhaustive communities that heuristically maximizes modularity,}

\begin{equation}
\mathcal{C}^{*} = \arg\max_{\mathcal{C}} \frac{1}{2m} \sum_{i,j} \left( w_{ij} - \frac{k_i k_j}{2m} \right) \delta(c_i, c_j)
\label{eq:modularity}
\end{equation}

\noindent \new{where $w_{ij}$ is the similarity weight of Eq.~\eqref{eq:weight}, $k_i = \sum_j w_{ij}$ is the weighted degree of vertex $v_i$, $m = \tfrac{1}{2}\sum_{i,j} w_{ij}$ is the total edge weight of the graph, $c_i$ denotes the community to which $v_i$ is assigned, and $\delta(c_i, c_j)$ is the Kronecker delta, equal to $1$ when $c_i = c_j$ and $0$ otherwise. The bracketed term contrasts the observed weight between $v_i$ and $v_j$ against the weight $k_i k_j / 2m$ expected under a null model that preserves degrees but rewires edges at random, so a partition scores highly only when its communities are denser than chance would explain. As noted in Section~\ref{sec:bg-louvain}, this maximization is NP-hard~\cite{brandes2008modularity} and Louvain solves it greedily; $\mathcal{C}^{*}$ should be read as the partition Louvain returns, not as a global optimum.}

\new{An alternative worth addressing explicitly is balanced graph partitioning, for which fast high-quality multilevel algorithms exist~\cite{catalyurek2023partitioning}. Balanced partitioners minimize edge cut subject to a constraint that all blocks have roughly equal size, and they require the number of blocks to be specified in advance. Neither property fits this problem. The number of distinct peptides in a mass bucket is unknown a priori and is exactly what we are trying to recover, and the true cluster-size distribution in shotgun proteomics is strongly skewed, with a few abundant peptides accounting for very large clusters and a long tail of singletons. A balance constraint would actively penalize the correct answer. One concern that motivates balanced partitioning does carry over, however: a very large community makes bundling more expensive and produces a representative that is less discriminative, since majority over many members regresses \gfix{toward} a vector that is moderately close to everything. \gfix{Section~\ref{sec:representatives}} returns to this point.}

\subsection{Cluster Representatives}
\label{sec:representatives}

\new{The offline phase closes by turning each community into an object the online phase can compare against. A community identifier cannot serve this purpose, since the assignment rule in Eq.~\eqref{eq:hamming} computes a Hamming distance between two hypervectors whereas a community label is an integer rather than a point in $\{0,1\}^D$. HDLouvain therefore derives one representative hypervector per community by bundling its members,}

\begin{equation}
r_C[\ell] \;=\;
\begin{cases}
1, & \text{if } \sum_{s \in C} h_s[\ell] > \tfrac{|C|}{2},\\[2pt]
0, & \text{otherwise,}
\end{cases}
\qquad \ell = 1, \ldots, D,
\label{eq:bundle}
\end{equation}
\new{that is, the bitwise majority over the member hypervectors defined in Section~\ref{sec:bg-hdc}, with ties at $|C|$ even resolved in \gfix{favor} of $0$. Bundling costs one pass over the community, $O(|C| \cdot D)$ counter updates followed by an $O(D)$ thresholding sweep, and is performed exactly once per community at the end of the offline phase. Note that this step does \emph{not} enjoy the $32\times$ bit-packing speedup that Eq.~\eqref{eq:hamming} does, because accumulating a per-position majority requires unpacked counters rather than word-wide logic; it is nonetheless a single linear pass over the bootstrap set and is negligible against the quadratic distance-matrix term.}

\new{Eq.~\eqref{eq:bundle} \emph{is} a centroid computation, so calling HDLouvain centroid-free would overstate the case. What it avoids is the \emph{iterative recomputation} that $k$-means and HAC pipelines perform, where every insertion invalidates a prototype. Here a representative is computed once from the Louvain partition and frozen, so per-spectrum online cost stays $O(|\mathcal{R}_b| \cdot D/32)$\final{, set by the active representatives in the bucket rather than by the spectra already ingested}. The cost is that a frozen representative cannot track membership drift over a long acquisition, which periodic re-bundling would fix at some throughput expense. A $k$-means baseline that also froze its centroids would not pay maintenance costs either, so what HDLouvain retains over such a baseline is cluster \emph{discovery}.} 

% \kevin{I suggest we add: Exploring coarse, periodic update of the centroids is an interesting direction for future work.}

\subsection{Online Incremental Assignment}
\label{sec:online}

\begin{figure}
\centering
\includegraphics[scale=0.35]{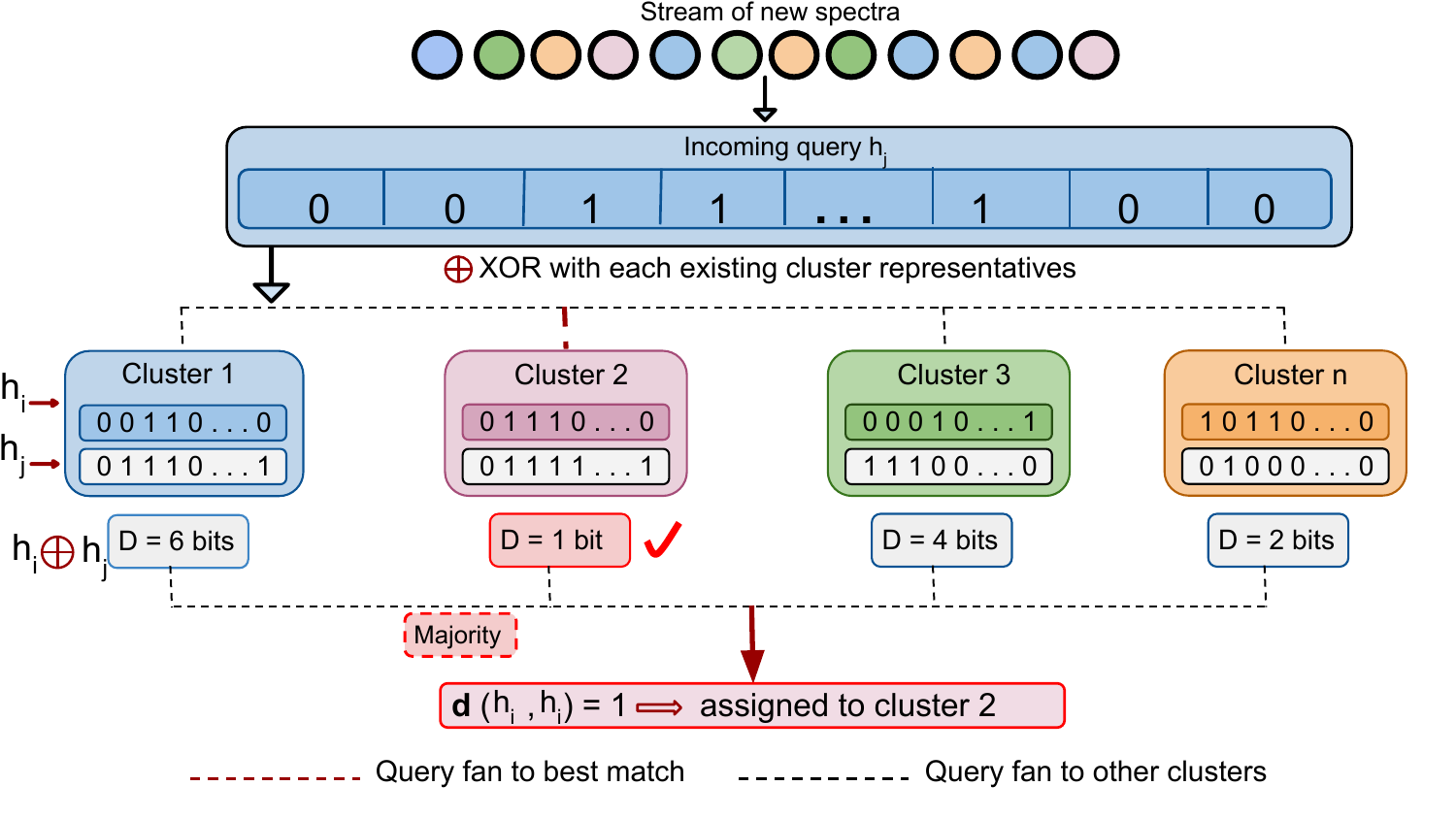}
\vspace{-0.25in}
      \caption{\textbf{Incremental assignment via Hamming distance.} \new{Within the bucket given by Eq.~\eqref{eq:bucket},} the query is XOR'd against the representative of \new{each cluster active in that bucket} and the population counts give the Hamming distances. It joins the nearest (here Cluster 2, at 1 bit), making affiliation a bit-parallel lookup rather than an all-pairs comparison. \new{Other buckets are never touched.}}
\label{fig:methodology}
\vspace{-0.15in}
\end{figure}

\new{Each arriving spectrum is processed in three constant-cost steps. It is first routed to a mass bucket by evaluating Eq.~\eqref{eq:bucket} on the precursor $m/z$ and charge in its own MGF record, which requires no comparison against stored data. It is then preprocessed and encoded exactly as in Section~\ref{sec:preprocessing}. Finally it is compared against $\mathcal{R}_b$, the representatives active in that bucket, and only those; representatives in other buckets are never touched, which is what the complexity analysis in Section~\ref{sec:complexity} assumes and what \textcolor{black}{Fig.~\ref{fig:methodology}} depicts within a single bucket.}

Both the incoming spectrum and every community representative are binary hypervectors in $\{0,1\}^D$, so the comparison is the normalized Hamming distance of Eq.~\eqref{eq:hamming}. The GPU implementation evaluates it with the CUDA integer intrinsics \texttt{XOR} and \texttt{popc}. By operating on bit-packed hypervectors, where $D$ bits are packed into $\lceil D/32 \rceil$ 32-bit words, the cost of each distance evaluation is reduced from $\mathcal{O}(D)$ to $\mathcal{O}(D/32)$, a 32-fold reduction that lets HDLouvain scale to large incremental workloads. \new{The representative set of a bucket is held as a contiguous bit-packed array, so a query is evaluated against all of $\mathcal{R}_b$ in one kernel launch with a parallel reduction for the $\arg\min$.}

\new{\textbf{The threshold $\tau$.} A spectrum joins its nearest representative
if the distance is at most $\tau$, and otherwise opens a new cluster. By
Eq.~\eqref{eq:hamming}\gfix{, $\tau$ is} a normalized Hamming distance in $[0,1]$, the
fraction of the $D$ bits by which a spectrum may differ from a representative and
still count as the same peptide. It is the only parameter of the online phase and
sets the cluster count, since too tight a value fragments clusters and inflates
$|\mathcal{R}_b|$ while too loose a value merges distinct peptides.}

\subsection{Complexity Analysis}
\label{sec:complexity}
\new{Let $n$ be the number of spectra, $B$ the number of non-empty mass buckets, $n_b$ the occupancy of bucket $b$, and $\mathcal{R}_b$ its active representatives.}

\new{The offline phase is dominated by the per-bucket distance matrices, $O\!\left(\sum_b n_b^2 \cdot D/32\right)$, still quadratic but in $n_b$ rather than $n$. For roughly balanced occupancy this is $O(n^2 D / (32B))$, a factor-$B$ reduction over global all-pairs comparison. Since Eq.~\eqref{eq:bucket} discretizes at the $1.0005$\,Da peptide mass spacing and tryptic precursor masses span roughly $10^2$ to $10^4$\,Da, $B$ is on the order of $10^3$--$10^4$, so the reduction is three to four orders of magnitude. Selecting the $k$ nearest \gfix{neighbors} costs $O(\sum_b n_b^2)$, one $O(n_b)$ selection per row, which the distance-matrix term dominates by $D/32$. Louvain contributes $O(kn)$, near-linear in the $O(k n_b)$ edges of each bucket graph, bundling one linear pass over the bootstrap set, and sorting by precursor $m/z$ an $O(n \log n)$ term that is not the bottleneck.}

\new{Locality-sensitive hashing, as in msCRUSH~\cite{wang2019mscrush}, and approximate \gfix{nearest-neighbor} indexing, as in Falcon~\cite{bittremieux2021falcon}, attack the same quadratic term more aggressively, reaching sub-quadratic \gfix{behavior} where bucketing only divides the constant. Bucketing is the cheaper and more conservative option, exact within a bucket, free of hash collisions and recall loss, and costing one arithmetic expression per spectrum with no index to build or maintain online. The two are composable, since nothing in the Louvain-then-bundle-then-assign core assumes how a bucket's candidate set was produced, and replacing exact intra-bucket distances with an LSH or ANN generator is the clearest route to removing the residual quadratic term.}

\subsection{Design Rationale}
\label{sec:rationale}

HDLouvain's core advantage is decoupling cluster \emph{discovery} from cluster \emph{assignment}. Louvain performs discovery once on the bootstrap fraction, producing high-modularity partitions that act as structural anchors, after which every arriving spectrum is assigned by bit-parallel XOR and popcount against the representatives, at a cost set by $D$ and $|\mathcal{R}_b|$.

The design therefore avoids full-dataset re-clustering, using the initial partition as a structural prior instead of recomputing from scratch, which mitigates the drift that affects purely online methods such as streaming DBSCAN or incremental HAC. The trade-off is deliberate, a bounded fidelity difference against a runtime reduction that makes high-throughput analysis tractable, and \new{Section~\ref{sec:sensitivity} shows it is monotonic in $\alpha$, so it can be \gfix{dialed} to a target rather than accepted as a fixed penalty.} Spectra that match no representative under $\tau$ open new clusters, so the partition still grows with the stream without revisiting the offline step.

\section{Evaluation}
\label{sec:eval}

We evaluate HDLouvain along three axes: clustering runtime against the static Louvain baseline (Section~\ref{sec:performance}), clustering quality on synthetic networks with LFR benchmark ground truth (Section~\ref{sec:accuracy}), and sensitivity to the offline/online split ratio $\alpha$ (Section~\ref{sec:sensitivity}).

\subsection{Datasets and Experimental Setup}

We evaluate HDLouvain on five public proteomics datasets spanning more than an order of magnitude in scale, from roughly $9.9\times 10^{5}$ to $2.1\times 10^{7}$ MS/MS spectra (11\,GB to 141\,GB on disk), and covering human kidney cell lysates (PXD001468, PXD001197), HeLa proteins (PXD003258), HEK293 cell lysates (PXD001511), and a draft map of the human proteome (PXD000561). All raw files were obtained from the PRIDE repository~\cite{perezriverol2019pride} and converted to MGF. \textcolor{black}{Table~\ref{tab:datasets}} \gfix{summarizes} their key properties.

\begin{table}[t]
\centering
\caption{Mass-spectrometry datasets used in our evaluation. All
datasets are obtained from the PRIDE repository~\cite{perezriverol2019pride}.}
\vspace{-0.05in}
\label{tab:datasets}
\begin{tabular}{|l|l|r|r|}
\hline
\textbf{PRIDE ID} & \textbf{Sample type} & \textbf{\#Spectra} & \textbf{Size} \\
\hline\hline
PXD001197~\cite{roos2016sil1}             & Kidney cell          & 992,847  & 25\,GB   \\
\hline
PXD001468~\cite{chick2015masstolerant}    & Kidney cell          & 1,022,004  & 11\,GB  \\
\hline
PXD001511~\cite{glatter2015comparison}    & HEK293 cell          & 3,594,140  & 87\,GB   \\
\hline
PXD003258~\cite{boisvert2012quantitative} & HeLa proteins        & 3,700,371  & 54\,GB   \\
\hline
PXD000561~\cite{kim2014draft}             & Human proteome       & 21,128,912 & 141\,GB  \\
\hline
\end{tabular}
\vspace{-0.15in}
\end{table}

Experiments are run on a workstation with an Intel Core i7-11700K CPU, 64\,GB DDR4, and an NVIDIA RTX 4090 with 24\,GB GDDR6X, under Ubuntu 20.04 LTS. HDLouvain is implemented in Python 3.8 in an \texttt{nvidia/cuda:11.8} container, using NumPy 1.21, SciPy 1.7, scikit-learn 1.0 and Numba 0.55 on the CPU, and CuPy 10.6, cuDF 22.8 and cuML 22.8 (CUDA 11.8) for GPU-accelerated $k$-NN construction, Hamming-distance evaluation and assignment. \new{The static baseline runs identical preprocessing, encoding, bucketing and graph-construction code on 100\% of the spectra, so the reported speedups isolate the offline/online decomposition rather than implementation differences.}

\subsection{Runtime Evaluation}
\label{sec:performance}

\textcolor{black}{Table~\ref{tab:louvain-clustering-time}} reports mean clustering runtime for static Louvain and HDLouvain across the five datasets at the balanced 50/50 split. Static Louvain scales steeply with the number of spectra, from roughly 34\,s on PXD001197 (0.99\,M spectra) to over 10{,}000\,s on PXD000561 (21.13\,M), an almost \gfix{three-order-of-magnitude} increase reflecting the super-linear cost of graph construction and modularity optimization. Within HDLouvain, the offline phase increasingly dominates the online one as spectra grow. The two smallest workloads run at near parity, with offline-to-online ratios of 0.96$\times$ and 0.97$\times$ where fixed setup overheads are a comparable share of both, but on PXD001511 the offline phase takes 83.21\,s against 44.77\,s online, a 1.9$\times$ asymmetry that reaches 2.4$\times$ on PXD003258 and nearly 4$\times$ on PXD000561. At scale the bit-parallel XOR/popcount kernel therefore costs far less per spectrum than graph construction and community detection, validating the choice to confine Louvain to a one-time bootstrap.

\begin{table*}[!t]
\centering
\caption{Clustering runtime of static Louvain versus HDLouvain across MS proteomics datasets (mean of 5 runs, seconds; $\alpha = 0.5$). \textit{Static}: full-dataset Louvain baseline; \textit{HDLouvain-Offline}: bootstrap phase, running Louvain on the initial 50\% of the stream; \textit{HDLouvain-Online}: incremental phase, assigning the remaining 50\% via GPU-accelerated Hamming distance; \textit{HDLouvain}: end-to-end runtime (Offline + Online); \textit{Speedup}: ratio of static Louvain runtime to end-to-end HDLouvain runtime.}
\vspace{-0.05in}
\label{tab:louvain-clustering-time}
\begin{tabular}{|l|r|r|r|r|r|r|}
\hline
\textbf{Dataset} & \textbf{\#Spectra} & \textbf{Static} & \makecell{\textbf{HDLouvain}\\\textbf{(Offline)}} & \makecell{\textbf{HDLouvain}\\\textbf{(Online)}} & \makecell{\textbf{HDLouvain}\\\textbf{(Offline + Online)}} & \textbf{Speedup} \\
\hline\hline
PXD001197 &    992{,}847 &    33.51\,s &   16.19\,s &  16.85\,s &   33.04\,s & 1.01$\times$ \\
\hline
PXD001468 &  1{,}022{,}004 &    37.77\,s &   17.17\,s &  17.67\,s &   34.84\,s & 1.08$\times$ \\
\hline
PXD001511 &  3{,}594{,}140 &   278.03\,s &   83.21\,s &  44.77\,s &  127.98\,s & 2.17$\times$ \\
\hline
PXD003258 &  3{,}700{,}371 &   477.65\,s &  132.44\,s &  55.64\,s &  188.08\,s & 2.54$\times$ \\
\hline
PXD000561 & 21{,}128{,}912 & 10103.80\,s & 2514.47\,s & 634.55\,s & 3149.02\,s & 3.21$\times$ \\
\hline
\end{tabular}
\vspace{-0.2 in}
\end{table*}

The speedup column follows the same trend, rising monotonically from 1.01$\times$ on the smallest dataset to 3.21$\times$ on the largest, with every dataset at or above break-even. The two datasets near 1\,M spectra sit essentially at parity (1.01$\times$ and 1.08$\times$), since the fixed setup costs of the hybrid pipeline, from per-bucket graph construction to GPU launch overhead, are not yet amortized, and static Louvain already finishes in under a minute there, leaving little headroom to recoup them. PXD001511 marks the inflection point, where the speedup jumps to 2.17$\times$ and then grows to 2.54$\times$ and 3.21$\times$. We therefore recommend HDLouvain for multi-million-spectra workloads, precisely the regime that motivates the framework.

\subsection{Clustering Quality Evaluation}
\label{sec:accuracy}

\begin{figure}[t]
    \centering
    \includegraphics[width=0.82\linewidth]{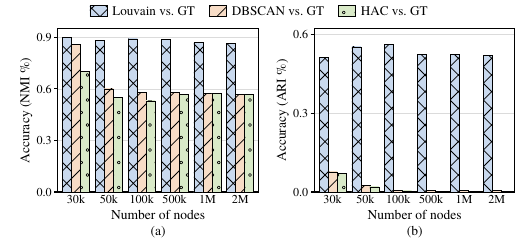}
    \vspace{-0.05in}
    \caption{Comparison between the LFR benchmark ground truth and clustering results: (a) NMI-based accuracy of Louvain, DBSCAN, and HAC against the ground-truth communities; (b) ARI-based accuracy of the same three methods against the LFR ground-truth communities.}
    \label{fig:accuracy}
    \vspace{-0.15in}
\end{figure}

\begin{figure}[t]
    \centering
    \includegraphics[width=0.82\linewidth]{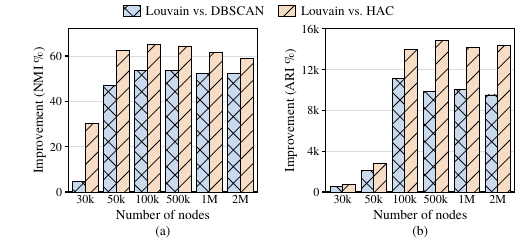}
    \vspace{-0.05in}
   \caption{Relative improvement of Louvain over DBSCAN and HAC on the LFR benchmark: (a) NMI-based improvement; (b) ARI-based improvement.}
    \label{fig:improvement}
    \vspace{-0.15in}
\end{figure}

\new{This section evaluates the \emph{community-detection stage} of HDLouvain, the offline component that supplies the structural prior for everything downstream, against the two clustering families it replaces.} We compare Louvain, DBSCAN, and HAC on synthetic networks with planted communities from the Lancichinetti--Fortunato--Radicchi (LFR) benchmark generator.
 
The LFR benchmark~\cite{lancichinetti2008benchmark} plants ground-truth communities in networks with power-law degree and community-size distributions, making it more demanding than the Girvan--Newman planted partition model~\cite{girvan2002community}. \new{We constructed networks at six scales, $N \in \{30\text{k},\, 50\text{k},\, 100\text{k},\, 500\text{k},\, 1\text{M},\, 2\text{M}\}$ nodes, using} degree exponent \final{$\gamma_1 = 2$}, community size exponent \final{$\gamma_2 = 1$}, average degree $\langle k \rangle = 20$, and maximum degree $k_{\text{max}} = 50$. \new{The mixing parameter was held at 0.6, since sweeping it is orthogonal to the scaling study reported here.} Louvain runs directly on the graph, while DBSCAN and HAC need a metric space, so we learn 128-dimensional node2vec embeddings~\cite{grover2016node2vec} ($p = 1$, $q = 1$, walk length $= 80$) and apply both to those. \new{All values are means over ten \gfix{realizations} per configuration.} Quality is assessed by \gfix{Normalized} NMI and ARI.

\begin{figure}[t]
    \centering
    \includegraphics[width=0.8\linewidth]{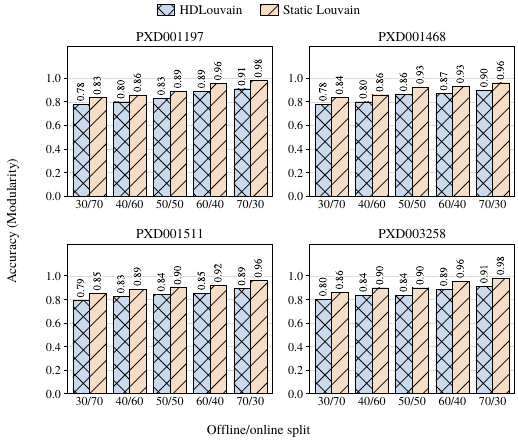}
    \vspace{-0.05in}
      \caption{\gnew{HDLouvain versus static Louvain as the split is swept. Modularity
rises for both, HDLouvain from 0.78--0.80 to 0.89--0.91 and static Louvain from
0.83--0.86 to 0.96--0.98, with the gap holding near 0.06 at every split.}}
    \label{fig:quality-splits}
    \vspace{-0.15in}
\end{figure}

 \begin{figure}[t]
    \centering
    \includegraphics[width=0.8\linewidth]{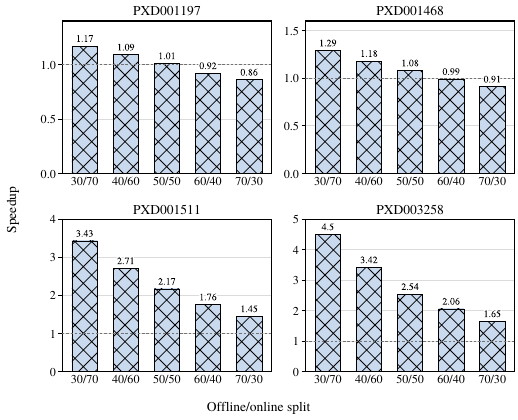}
    \vspace{-0.05in}
    \caption{Speedup of HDLouvain over static Louvain on four MS datasets at offline/online splits from 30/70 to 70/30. It falls monotonically with the offline fraction and grows with dataset size, peaking at 4.5$\times$ on PXD003258; the two smaller datasets fall below break-even at high offline fractions.}
    \label{fig:sensitivity-speedup}
    \vspace{-0.15in}
\end{figure}

Louvain aligns substantially better with ground truth at every scale, as shown in \gfix{Fig.~\ref{fig:accuracy}}. Under NMI it scores 0.86--0.93, while DBSCAN collapses from 0.86 to roughly 0.57 beyond 30k nodes and HAC falls from 0.70 to roughly 0.57 at 2M. Under ARI, which corrects for chance over all pairwise assignments and so penalizes even modest boundary error, Louvain retains 0.33--0.61 whereas DBSCAN and HAC drop to 0.002--0.009 at 100k nodes and above, three orders of magnitude lower and effectively indistinguishable from random assignment. \textcolor{black}{Fig.~\ref{fig:improvement}} presents the corresponding percentage improvements.
 
We also measure clustering quality on four real PRIDE datasets, using \gfix{modularity} across five splits from 30/70 to 70/30. \gnew{Fig.~\ref{fig:quality-splits} reports the outcome.} Both methods improve monotonically with the offline fraction, HDLouvain from
0.78--0.80 at 30/70 to 0.89--0.91 at 70/30 and Static Louvain from 0.83--0.86
to 0.96--0.98, since more initial topology gives the optimizer more to work with
and static recomputation always sees the complete graph. The gap is bounded across every dataset and split, with a maximum degradation
of 8\%, and it buys a $4.5\times$ speedup on the largest dataset in the sweep, obtained by \final{assigning each arriving spectrum by a bit-parallel Hamming lookup against the frozen bucket representatives} rather than recomputing the full graph.

\subsection{Sensitivity to the Offline/Online Split Ratio}
\label{sec:sensitivity}

To establish that HDLouvain's behavior is not tied to a single split, we sweep the bootstrap fraction $\alpha$, the share of the stream handled offline, over $\{0.3, \ldots, 0.7\}$ for runtime\final{ and over $[0.2, 0.9]$ for quality}, on the four smaller datasets.

\textcolor{black}{Fig.~\ref{fig:sensitivity-speedup}} reports speedup over the static baseline. It falls monotonically with $\alpha$ on every dataset, peaks at 30/70, and grows with dataset size to $4.5\times$ on PXD003258 and $3.43\times$ on PXD001511. The two smaller datasets reach only 1.17$\times$ and 1.29$\times$ there and fall below break-even as $\alpha$ approaches 0.7.

\begin{figure}[t]
    \centering
\includegraphics[width=0.8\linewidth]{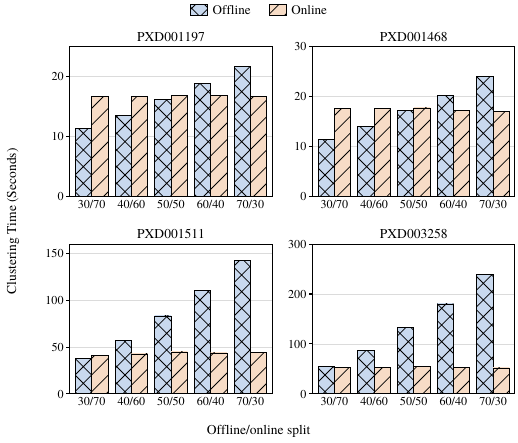}
    \vspace{-0.05in}
    \caption{Runtime of HDLouvain decomposed into offline and online phases on four MS datasets at offline/online splits from 30/70 to 70/30. The bootstrap phase grows with the offline fraction while the online phase stays nearly flat, reflecting the lower per-spectrum cost of the HDC kernel.}
    \label{fig:sensitivity-time}
    \vspace{-0.15in}
\end{figure}

\textcolor{black}{Fig.~\ref{fig:sensitivity-time}} decomposes runtime into its two phases. The offline component grows monotonically with $\alpha$, since a larger bootstrap sends more spectra through the $k$-NN graph and modularity optimization, which is why 70/30 approaches the cost of full static Louvain. The online component stays nearly flat, since the kernel's per-spectrum cost is low enough that the drop in volume barely registers.

\begin{figure}[t]
    \centering
    \includegraphics[width=0.8\linewidth]{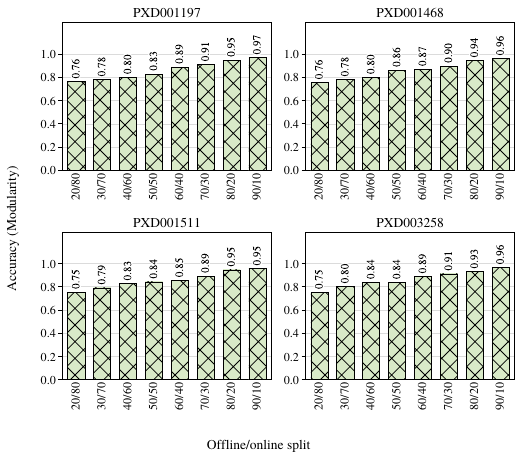}
    \vspace{-0.05in}
       \caption{\gnew{HDLouvain quality over a wider sweep. Modularity rises
monotonically with the offline fraction on all four datasets, from 0.75--0.76 at
20/80 to 0.95--0.97 at 90/10, within 0.02 of static Louvain at 70/30 in
Fig.~\ref{fig:quality-splits}.}}
    \label{fig:sensitivity-accuracy}
    \vspace{-0.2in}
\end{figure}

\new{\textcolor{black}{Fig.~\ref{fig:sensitivity-accuracy}} shows the quality side
of the trade-off, which is why we present $\alpha$ as a tunable knob rather than a
single headline configuration.} \gnew{Extending the sweep of
Fig.~\ref{fig:quality-splits} in both directions, from a 20\% to a 90\% offline
bootstrap, modularity rises monotonically with $\alpha$ on all four datasets, from
0.75--0.76 at 20/80 to 0.95--0.97 at 90/10. Two features of the curve matter for
choosing $\alpha$. First, even the most aggressive 20/80 split holds modularity at
0.75, so the penalty for running the pipeline at a small bootstrap is bounded
rather than catastrophic. Second, the curve keeps rising past the 70/30 endpoint
of Fig.~\ref{fig:quality-splits}, and by 90/10 it sits within 0.02 of the
0.96--0.98 that static Louvain attains there, so a large enough bootstrap all but
removes the residual cost of never revisiting the offline partition.}

\new{Read with \textcolor{black}{Fig.~\ref{fig:sensitivity-speedup}}, the two
curves move in opposite directions, so $\alpha$ should follow from the
application's quality budget. Where the partition must come} \gnew{as close as
possible to} \new{a full static analysis, $\alpha \ge 0.7$ still returns speedups of 
1.45$\times$ and 1.65$\times$ on the two multi-million-spectra datasets while}
\gnew{closing to within about 7\% of static-Louvain modularity;} \new{the smaller
datasets fall below break-even there and are better served by the static
pipeline. When a coarser partition suffices, for instance for redundancy
reduction ahead of a database search, $\alpha \in [0.3, 0.5]$ delivers
2.17$\times$ to 4.5$\times$ speedups.}  

% \kevin{I suggest to add: Larger datasets might allow for smaller bootstrap fractions at a given quality budget, assuming the bootstrap data are representative of the remaining data elements. This is an area for future work.}

\section{Related Work}
\textbf{\textit{Clustering of Tandem Mass Spectra}}. MS-Clustering~\cite{frank2008msclustering} and PRIDE Cluster~\cite{griss2013prideclustering} scaled greedy and hierarchical strategies to repository size at the cost of representative drift and periodic recomputation. Later work attacked the bottleneck, with MaRaCluster~\cite{the2016maracluster} replacing cosine similarity by a fragment-rarity statistic, msCRUSH~\cite{wang2019mscrush} using locality-sensitive hashing, and Falcon~\cite{bittremieux2021falcon} using approximate nearest-neighbor indexing to reach billions of spectra, though none dominates on runtime, sensitivity, and purity together~\cite{rieder2017comparison}. ClusterSheep~\cite{to2021clustersheep} ported HAC to the GPU but inherited its quadratic memory, and Spectroscape~\cite{wu2023spectroscape} reframed the problem as real-time archive query. None of these ingests incrementally without paying much of the offline cost per batch. \new{Wang et al.~\cite{wang2026incremental} come closest, clustering at batch granularity by pooling spectra to propagate density structure across batches, scaling to 368\,M spectra; HDLouvain assigns at spectrum granularity against frozen representatives, re-deriving nothing per batch. It also complements approximate search, since msCRUSH and Falcon cut the candidate count whereas HDLouvain cuts how often discovery runs at all (Section~\ref{sec:complexity}).}

\textit{\textbf{HDC-Based Clustering}.}
Hyperdimensional computing~\cite{kanerva2009hdc, heddes2024hdc} suits GPUs, \gfix{FPGAs, and} processing-in-memory substrates but is sensitive to prototype initialization, since random seed hypervectors yield unstable partitions that assignment efficiency cannot recover from~\cite{ge2024robust}. Its MS applications, HyperSpec~\cite{xu2023hyperspec}, SpecHD~\cite{pinge2024spechd}, RapidOMS~\cite{pinge2024rapidoms}\gfix{,} and SmartMS~\cite{ponzina2025smartms}, are all static and batch-oriented. HDLouvain inherits the bit-parallel kernel, replaces random prototypes with bundled Louvain communities, and lifts the static restriction by treating those bundles as long-lived representatives.

\textit{\textbf{Graph Community Detection}.}
Louvain~\cite{blondel2008louvain} and its Leiden refinement~\cite{traag2019leiden} are batch procedures, which has motivated incremental variants. DynaMo~\cite{zhuang2019dynamo} re-runs local moves on changed edges, DF~Louvain~\cite{sahu2024dflouvain} parallelizes this with a dynamic frontier, and Wang et al.~\cite{wang2026realtime} reach sub-second latency on billion-scale graphs. All assume modest perturbations, failing under k-edge arrivals; HDLouvain freezes offline communities and delegates assignment to HDC similarity. \new{Balanced (hyper)graph partitioning~\cite{catalyurek2023partitioning} offers fast multilevel algorithms, but as Section~\ref{sec:communities} discusses, Its balance constraint and fixed block count are mismatched to a heavy-tailed, unknown-cluster-count distribution}

\section{Conclusions}
\new{We presented HDLouvain, which runs Louvain once on a bootstrap fraction 
$\alpha$, bundles each community into one binary representative, and assigns later spectra via bit-parallel Hamming lookup against bucket-local representatives }\final{, cost set by representative count, not stream length}. On five datasets spanning 0.99--21.13\,M spectra it reaches 3.21$\times$ over static Louvain at $\alpha = 0.5$\final{, 4.5$\times$ at $\alpha = 0.3$ on the four swept datasets, and 1.45--1.65$\times$ at $\alpha \ge 0.7$}.As future work, we plan to explore a periodic cycle of stair-step re-clustering, incremental assignment, and cleanup to bound cluster drift and improve clustering quality over long-running streams

 \section{Acknowledgment}
 This work was supported in part by PRISM, one of seven centers in JUMP 2.0, an SRC program sponsored by DARPA.

\bibliographystyle{IEEEtran}
\bibliography{refs}

\end{document}